\input epsf.tex
%
\def\inv{^{\raise.15ex\hbox{${\scriptscriptstyle -}$}\kern-.05em 1}}
\def\lbar{{\lower.35ex\hbox{$\mathchar'26$}\mkern-10mu\lambda}} 

%
%
%
%
\def\dsl{\,\raise.15ex\hbox{/}\mkern-13.5mu D} 
\def\delsl{\raise.15ex\hbox{/}\kern-.57em\partial}
\def\Ksl{\hbox{/\kern-.6000em\rm K}}
\def\Asl{\hbox{/\kern-.6500em \rm A}}
\def\Dsl{\hbox{/\kern-.6000em\rm D}} 
\def\Qsl{\hbox{/\kern-.6000em\rm Q}}
\def\gradsl{\hbox{/\kern-.6500em$\nabla$}}
%
%
\def\lspace{\ifx\answ\bigans{}\else\qquad\fi}
\def\lbspace{\ifx\answ\bigans{}\else\hskip-.2in\fi} 
%
%
\def\boxeqn#1{\vcenter{\vbox{\hrule\hbox{\vrule\kern3pt\vbox{\kern3pt
        \hbox{${\displaystyle #1}$}\kern3pt}\kern3pt\vrule}\hrule}}}
%
%
\def\mbox#1#2{\vcenter{\hrule \hbox{\vrule height#2in
\kern#1in \vrule} \hrule}}
%
%
%
%
   
 \def\CF{{\cal F}}  
   
  \def\CO{{\cal O}}

%
%
%
%
%

%

\def\bar#1{\overline{#1}}
\def\vev#1{\left\langle #1 \right\rangle}

\def\darr#1{\raise1.5ex\hbox{$\leftrightarrow$}\mkern-16.5mu #1}

%
%
\def\frac#1#2{{\textstyle{#1\over #2}}} 
%
%
%
%

%
%
%
%
 
%
%
\def\ltap{\ \raise.3ex\hbox{$<$\kern-.75em\lower1ex\hbox{$\sim$}}\ }
\def\gtap{\ \raise.3ex\hbox{$>$\kern-.75em\lower1ex\hbox{$\sim$}}\ }
\def\gl{\ \raise.5ex\hbox{$>$}\kern-.8em\lower.5ex\hbox{$<$}\ }
\def\roughly#1{\raise.3ex\hbox{$#1$\kern-.75em\lower1ex\hbox{$\sim$}}}
%
%
\def\ie{\hbox{\it i.e.}}        
        
\def\etal{\hbox{\it et al.}}

\def\np#1#2#3{{Nucl. Phys. } B{#1} (#2) #3}
\def\pl#1#2#3{{Phys. Lett. } {#1}B (#2) #3}
\def\prl#1#2#3{{Phys. Rev. Lett. } {#1} (#2) #3}
\def\physrev#1#2#3{{Phys. Rev. } {#1} (#2) #3}

\relax
%
%
%

%
%
%
\newbox\leftpage \newdimen\fullhsize \newdimen\hstitle \newdimen\hsbody
\tolerance=1000\hfuzz=2pt
\def\printertype{}
\def\qms{\def\printertype{qms: }
\ifx\answ\bigans\else\voffset=-.4truein\hoffset=.125truein\fi}
\def\bigans{b }
\message{ big or little (b/l)? }\read-1 to\answ
\ifx\answ\bigans\message{(This will come out unreduced.}
\magnification=1200\baselineskip=16pt plus 2pt minus 1pt
\hsbody=\hsize \hstitle=\hsize 
\else\message{(This will be reduced.} \let\lr=L
\magnification=1000\baselineskip=16pt plus 2pt minus 1pt
\voffset=-.31truein\vsize=7truein\hoffset=-.465truein
\hstitle=8truein\hsbody=4.75truein\fullhsize=10truein\hsize=\hsbody
\output={ 
  \almostshipout{\leftline{\vbox{\pagebody\makefootline}}}\advancepageno
}
\def\almostshipout#1{\if L\lr \count1=1 \message{[\the\count0.\the\count1]}
      \global\setbox\leftpage=#1 \global\let\lr=R
  \else \count1=2
    \shipout\vbox{
           \special{\printertype landscape}
      \hbox to\fullhsize{\box\leftpage\hfil#1}}  \global\let\lr=L\fi}
\fi
%
\catcode`\@=11 
\newcount\yearltd\yearltd=\year\advance\yearltd by -1900
%
%
%
%
%
%
%
%
\def\title#1{\nopagenumbers\hsize=\hsbody%
\centerline{\titlefont #1} \tenpoint \vskip .5in\pageno=0}
%
%

\def\mayer{\vbox{\sl\centerline{Department of Physics 0319}%
\centerline{University of California, San Diego}
\centerline{9500 Gilman Drive}
\centerline{La Jolla, CA 92093-0319}}}
%
%
%

\def\doe{\#DE-FG03-90ER40546}

\def\UCSD#1#2{\noindent#1\hfill #2%
\bigskip\supereject\global\hsize=\hsbody%
\footline={\hss\tenrm\folio\hss}}
%
%
%
\def\abstract#1{\centerline{\bf Abstract}\nobreak\medskip\nobreak\par #1}
%
%

\def\draftmode{\message{ DRAFTMODE }\def\draftdate{{\rm preliminary draft:
\number\month/\number\day/\number\yearltd\ \ \hourmin}}%
\headline={\hfil\draftdate}\writelabels\baselineskip=20pt plus 2pt minus 2pt
 {\count255=\time\divide\count255 by 60 \xdef\hourmin{\number\count255}
  \multiply\count255 by-60\advance\count255 by\time
  \xdef\hourmin{\hourmin:\ifnum\count255<10 0\fi\the\count255}}}
\def\nolabels{\def\wrlabel##1{}\def\eqlabel##1{}\def\reflabel##1{}}
\def\writelabels{\def\wrlabel##1{\leavevmode\vadjust{\rlap{\smash%
{\line{{\escapechar=` \hfill\rlap{\sevenrm\hskip.03in\string##1}}}}}}}%
\def\eqlabel##1{{\escapechar-1\rlap{\sevenrm\hskip.05in\string##1}}}%
\def\reflabel##1{\noexpand\llap{\noexpand\sevenrm\string\string\string##1}}}
\nolabels
%
\global\newcount\secno \global\secno=0
\global\newcount\meqno \global\meqno=1
\def\newsec#1{\global\advance\secno by1\message{(\the\secno. #1)}
\global\subsecno=0\xdef\secsym{\the\secno.}\global\meqno=1
\bigbreak\bigskip\noindent{\bf\the\secno. #1}\writetoca{{\secsym} {#1}}
\par\nobreak\medskip\nobreak}
\xdef\secsym{}
\global\newcount\subsecno \global\subsecno=0
\def\subsec#1{\global\advance\subsecno by1\message{(\secsym\the\subsecno. #1)}
\bigbreak\noindent{\it\secsym\the\subsecno. #1}\writetoca{\string\quad
{\secsym\the\subsecno.} {#1}}\par\nobreak\medskip\nobreak}
\def\appendix#1#2{\global\meqno=1\global\subsecno=0\xdef\secsym{\hbox{#1.}}
\bigbreak\bigskip\noindent{\bf Appendix #1. #2}\message{(#1. #2)}
\writetoca{Appendix {#1.} {#2}}\par\nobreak\medskip\nobreak}
%
%
\def\eqnn#1{\xdef #1{(\secsym\the\meqno)}\writedef{#1\leftbracket#1}%
\global\advance\meqno by1\wrlabel#1}
\def\eqna#1{\xdef #1##1{\hbox{$(\secsym\the\meqno##1)$}}
\writedef{#1\numbersign1\leftbracket#1{\numbersign1}}%
\global\advance\meqno by1\wrlabel{#1$\{\}$}}
\def\eqn#1#2{\xdef #1{(\secsym\the\meqno)}\writedef{#1\leftbracket#1}%
\global\advance\meqno by1$$#2\eqno#1\eqlabel#1$$}
%
\newskip\footskip\footskip14pt plus 1pt minus 1pt 
\def\f@@t{\baselineskip\footskip\bgroup\aftergroup\@foot\let\next}
\setbox\strutbox=\hbox{\vrule height9.5pt depth4.5pt width0pt}
\global\newcount\ftno \global\ftno=0
\def\foot{\global\advance\ftno by1\footnote{$^{\the\ftno}$}}
%
\newwrite\ftfile
\def\footend{\def\foot{\global\advance\ftno by1\chardef\wfile=\ftfile
$^{\the\ftno}$\ifnum\ftno=1\immediate\openout\ftfile=foots.tmp\fi%
\immediate\write\ftfile{\noexpand\smallskip%
\noexpand\item{f\the\ftno:\ }\pctsign}\findarg}%
\def\footatend{\vfill\eject\immediate\closeout\ftfile{\parindent=20pt
\centerline{\bf Footnotes}\nobreak\bigskip\input foots.tmp }}}
\def\footatend{}
%
%
\global\newcount\refno \global\refno=1
\newwrite\rfile
\def\ref{[\the\refno]\nref}
\def\nref#1{\xdef#1{[\the\refno]}\writedef{#1\leftbracket#1}%
\ifnum\refno=1\immediate\openout\rfile=refs.tmp\fi
\global\advance\refno by1\chardef\wfile=\rfile\immediate
\write\rfile{\noexpand\item{#1\ }\reflabel{#1\hskip.31in}\pctsign}\findarg}
\def\findarg#1#{\begingroup\obeylines\newlinechar=`\^^M\pass@rg}
{\obeylines\gdef\pass@rg#1{\writ@line\relax #1^^M\hbox{}^^M}%
\gdef\writ@line#1^^M{\expandafter\toks0\expandafter{\striprel@x #1}%
\edef\next{\the\toks0}\ifx\next\em@rk\let\next=\endgroup\else\ifx\next\empty%
\else\immediate\write\wfile{\the\toks0}\fi\let\next=\writ@line\fi\next\relax}}
\def\striprel@x#1{} \def\em@rk{\hbox{}}

\def\addref#1{\immediate\write\rfile{\noexpand\item{}#1}} 
\def\startrefs#1{\immediate\openout\rfile=refs.tmp\refno=#1}
\def\xref{\expandafter\xr@f}\def\xr@f[#1]{#1}
\def\refs#1{[\r@fs #1{\hbox{}}]}
\def\r@fs#1{\edef\next{#1}\ifx\next\em@rk\def\next{}\else
\ifx\next#1\xref #1\else#1\fi\let\next=\r@fs\fi\next}
%

%
\newwrite\ffile\global\newcount\figno \global\figno=1
\def\fig{fig.~\the\figno\nfig}
\def\nfig#1{\xdef#1{fig.~\the\figno}%
\writedef{#1\leftbracket fig.\noexpand~\the\figno}%
\ifnum\figno=1\immediate\openout\ffile=figs.tmp\fi\chardef\wfile=\ffile%
\immediate\write\ffile{\noexpand\medskip\noexpand\item{Fig.\ \the\figno. }
\reflabel{#1\hskip.55in}\pctsign}\global\advance\figno by1\findarg}
\def\xfig{\expandafter\xf@g}\def\xf@g fig.\penalty\@M\ {}
\def\figs#1{figs.~\f@gs #1{\hbox{}}}
\def\f@gs#1{\edef\next{#1}\ifx\next\em@rk\def\next{}\else
\ifx\next#1\xfig #1\else#1\fi\let\next=\f@gs\fi\next}
\newwrite\lfile
{\escapechar-1\xdef\pctsign{\string\%}\xdef\leftbracket{\string\{}
\xdef\rightbracket{\string\}}\xdef\numbersign{\string\#}}

\def\writestop{\def\writestoppt{\immediate\write\lfile{\string\pageno%
\the\pageno\string\startrefs\leftbracket\the\refno\rightbracket%
\string\def\string\secsym\leftbracket\secsym\rightbracket%
\string\secno\the\secno\string\meqno\the\meqno}\immediate\closeout\lfile}}
\def\writestoppt{}\def\writedef#1{}
\def\seclab#1{\xdef #1{\the\secno}\writedef{#1\leftbracket#1}\wrlabel{#1=#1}}
\def\subseclab#1{\xdef #1{\secsym\the\subsecno}%
\writedef{#1\leftbracket#1}\wrlabel{#1=#1}}
\newwrite\tfile \def\writetoca#1{}
\def\leaderfill{\leaders\hbox to 1em{\hss.\hss}\hfill}
\def\writetoc{\immediate\openout\tfile=toc.tmp
   \def\writetoca##1{{\edef\next{\write\tfile{\noindent ##1
   \string\leaderfill {\noexpand\number\pageno} \par}}\next}}}

%
\catcode`\@=12 
%
\font\titlerm=cmr10 scaled\magstep3 \font\titlerms=cmr7 scaled\magstep3
\font\titlermss=cmr5 scaled\magstep3 \font\titlei=cmmi10 scaled\magstep3
\font\titleis=cmmi7 scaled\magstep3 \font\titleiss=cmmi5 scaled\magstep3
\font\titlesy=cmsy10 scaled\magstep3 \font\titlesys=cmsy7 scaled\magstep3
\font\titlesyss=cmsy5 scaled\magstep3 \font\titleit=cmti10 scaled\magstep3
\skewchar\titlei='177 \skewchar\titleis='177 \skewchar\titleiss='177
\skewchar\titlesy='60 \skewchar\titlesys='60 \skewchar\titlesyss='60
\def\titlefont{\def\rm{\fam0\titlerm}
\textfont0=\titlerm \scriptfont0=\titlerms \scriptscriptfont0=\titlermss
\textfont1=\titlei \scriptfont1=\titleis \scriptscriptfont1=\titleiss
\textfont2=\titlesy \scriptfont2=\titlesys \scriptscriptfont2=\titlesyss
\textfont\itfam=\titleit \def\it{\fam\itfam\titleit} \rm}
\def\tenpoint{\def\rm{\fam0\tenrm}
\textfont0=\tenrm \scriptfont0=\sevenrm \scriptscriptfont0=\fiverm
\textfont1=\teni  \scriptfont1=\seveni  \scriptscriptfont1=\fivei
\textfont2=\tensy \scriptfont2=\sevensy \scriptscriptfont2=\fivesy
\textfont\itfam=\tenit \def\it{\fam\itfam\tenit}
\textfont\bffam=\tenbf \def\bf{\fam\bffam\tenbf} \rm}
%
%
\def\noblackbox{\overfullrule=0pt}
\hyphenation{anom-aly anom-alies coun-ter-term coun-ter-terms}
\relax
\ifx\answ\bigans
	\def\gapi{6.3pt}
	\def\gapii{12.8pt}
	\def\pag{{ }}
\else
	\def\gapi{3.75pt}
	\def\gapii{10.0pt}
	\def\pag{\goodbreak}
\fi

\def\caption#1{
	\centerline{\vbox{\baselineskip=12pt
	\vskip.15in\hsize=3.8in\noindent{#1}\vskip.1in }}}
\def\trip#1#2#3{\(\matrix{#1 \cr #2\cr #3\cr}\)}
\def\frac#1#2{{\textstyle{#1 \over #2}}}
\def\[{\left[}
\def\]{\right]}
\def\({\left(}
\def\){\right)}

\def\doe{\#DOE-FG03-90ER40546}

\def\pyidk{PHY-9057135}
\nref\sut{W.M. Fairbairn, T. Fulton and W.H. Klink, Jour. Math. Phys. 5 (1964)
1038}
\noblackbox
\vskip 1.in
\centerline{{\titlefont Flavor Unification and Discrete}}
\vskip.2in
\title{  Nonabelian Symmetries }
\bigskip\bigskip
\centerline{David B. Kaplan\footnote{$^\dagger$}{Alfred
P. Sloan Fellow and NSF Presidential Young Investigator. }
and Martin Schmaltz}
\bigskip\mayer
\bigskip
\vfill\abstract{
Grand unified theories with fermions transforming as irreducible
representations of a discrete nonabelian flavor symmetry can lead to
realistic fermion masses, without requiring small fundamental parameters.   We
construct a specific example of a supersymmetric GUT based on the flavor
symmetry $\Delta(75)$ --- a subgroup of $SU(3)$ --- which can explain the
observed quark and lepton masses and mixing angles. The model predicts
$\tan\beta \simeq 2-5$ and gives a $\tau$ neutrino mass $m_\nu\simeq M_p/G_F
M_{GUT}^2 = 10$ eV, with other neutrino masses much lighter. Combined
constraints of light quark masses and perturbative unification place flavor
symmetry breaking near the GUT scale;  it may be possible to probe these
extremely high energies by continuing the search for flavor changing neutral
currents. }
\vfill\UCSD{UCSD/PTH 93-30, 
\ hep-ph/9311281
}{October 1993}

\newsec{Introduction}

 Particle physics seems to be at a stage similar to chemistry before Mendeleev,
or spectroscopy before Balmer---we are confronted with apparent patterns in
quark and lepton masses and mixing angles, yet have no compelling explanation
for them. It is likely that the difficulty is due to several simultaneous
effects contributing to the observed mass relations. These effects could
include radiative corrections in scaling from short distances, Clebsch factors
from gauge groups, mass matrix ``textures'' and Clebsch factors from flavor
symmetry groups, flavor symmetry breaking vacuum alignment, and higher
dimension operators induced by quantum gravity.  Aside from the observed
masses, the only experimental evidence we have to guide us is the absence of
flavor changing neutral currents (FCNC). In order to make headway in the face
of such ignorance it is necessary to have esthetic prejudices for guidance; in
this letter we adopt several. The first prejudice is that the fundamental
theory not contain parameters less than $\CO(10^{-1})$. The second is the
principle of ``flavor democracy'' \ref\dem{H. Georgi, A.V. Manohar and A.E.
Nelson,\pl{126}{1983}{169}
}, namely that all fermions with identical gauge charges have the same or
similar short distance interactions, with the observed diversity in masses
arising from dynamics.   Thirdly, we only consider theories where the gauge
interactions are unifiable, in order to adopt the successes in explaining the
equality of the proton and positron charges, as well as
predicting $\sin^2\theta_w$ and the relations between quark and lepton masses
\nref\guts{H. Georgi and S. Glashow, \prl{32}{1974}{438}; H. Georgi, H.R. Quinn
and S. Weinberg, \prl{33}{1974}{451};
A.J. Buras \etal, \np{135}{1978}{66};
S. Dimopoulos, S. Raby and F. Wilczek, \physrev{D24}{1981}{1681};
S. Dimopoulos and H. Georgi, \np{193}{1981}{150}}
\nref\bbo{V. Barger, M.S. Berger, P. Ohmann,
\physrev{D47}{1993}{109}}\nref\cpr{ H. Arason, D.J. Casta{\~n}o, E. J. Piard,
and P. Ramond, \physrev{D47}{1993}{232}; D.J. Casta{\~n}o, E.J. Piard, P.
Ramond, UFIFT-HEP-93-18 (1993) hep-ph-9308335}\refs{\guts-\cpr}.

As we will show, these three prejudices naturally lead us to consider theories
with nonabelian discrete flavor symmetries.  Such symmetries allow us to
understand many features of the quark and lepton masses, such as why the down
type quarks are lighter than up quarks in all but the first generation, and why
the Cabbibo angle is much larger than the other KM angles. The type of theories
we consider typically require flavor symmetry breaking to be near the GUT scale
and offer the tantalizing prospect of probing GUT-scale physics through
searches for flavor changing neutral currents (FCNC).  They also suggest that
the neutrinos are massive, with the tau neutrino mass naturally in the range
favored for dark matter.

The principles we adopt force us to think carefully about flavor symmetries. In
order to explain in a natural way a small mass ratio such as $m_e/ m_t \sim
3\times 10^{-6}$ in terms of parameters $\epsilon\sim 10^{-1}$, we must assume
that the mass ratios arise as high powers of $\epsilon$. These powers of
$\epsilon$ can arise naturally if $\epsilon$  measures mixing between ordinary
fermions and massive exotic fermions through soft flavor symmetry breaking
\ref\frog{C.D. Frogatt and H.B. Nielsen, \np{147}{1979}{277}}. Then $\epsilon
\sim g\langle X\rangle/M$, where $g$ is a coupling constant, $\langle X\rangle$
is a soft flavor symmetry breaking parameter, and $M$ is the heavy fermion
mass. The invariant tensors of the broken flavor symmetry group and pattern of
symmetry breaking naturally impose a texture on the effective Yukawa couplings
of the low energy theory \foot{The has been much recent interest in
investigating acceptable and predictive mass matrix textures; see, for example
\nref\hetal{
S. Dimopoulos, L.J. Hall and Stuart Raby, \prl{68}{1992}{1984};
G. W. Anderson \etal, \physrev{D47}{1993}{3702};
L.J. Hall and A. Rasin, \pl{315}{1993}{164} }\nref\rametal{P. Ramond, R.G.
Roberts and G.G. Ross, \np{406}{1993}{19}}\refs{\hetal,\rametal}.}.
The goal then is to find models which lead to a phenomenologically acceptable
texture. Most previous work in this direction has focused on Abelian flavor
symmetries ($U(1)$ or $Z_N$) which allow one to ``dial'' the fermion mass
matrices by judiciously choosing the charges for each fermion; for a recent
example consistent with current phenomenology, see ref. \ref\nati{M. Leurer, Y.
Nir and N. Seiberg, \np{398}{1993}{319};
 RU-93-43 (1993), hep-ph-9310320
}.
Pouliot and Seiberg have also constructed a nonabelian example of such models,
based on $O(2)\times U(1)$ \ref\nonab{P. Pouliot and N. Seiberg, Rutgers
preprint RU-93-39, hep-ph-9308363}, with the quarks in reducible
representations. Since  all of these models have quarks and leptons in
reducible flavor representations, the different generations are distinguished
by their flavor charges and have different interactions. However, this is  not
compatible with our goal of flavor democracy, which can only be achieved  by
putting all particles of like gauge charge in irreducible flavor
representations.  Furthermore, existing approaches do not lend themselves
readily to a unification of gauge forces.

In order to unify the three families into irreducible flavor triplets we are
compelled to search for a nonabelian flavor symmetry $G_f$ with one or more
three dimensional representations. For continuous symmetries, this only allows
groups with at least one factor of $SO(3)$, $SU(2)$ or $SU(3)$. A further
restriction is found by considering the top quark, whose mass must arise at
$\CO(\epsilon^0)$, if it is to have perturbative interactions.  Thus the
operator
\eqn\upmss{QU^cH_u}
 must be a $G_f$ invariant and lead to a rank one mass matrix.  If $Q$ and
$U^c$ are to be triplets of $G_f$, and $H_u$ is some irreducible
representation, then we can rule out the possibilities $G_f=SU(2)$ and
$G_f=SO(3)$ --- for those groups the operator \upmss\ yields a mass matrix that
is  either the unit matrix or traceless, and hence at least rank two. Similar
reasoning excludes $G_f=SU(3)$ unless $Q$ and $U$ transform as $3$'s and $H_u$
as a $\bar 6$ with  $\vev{H_u} = v\delta_{33}$.  A semisimple group such as
 $G_f= SU(3)\times SU(3)$ with $Q=(3,1)$, $U=(1,3)$, $H_u=(\bar 3,\bar 3)$ is a
possibility, as are groups with more factors.

 The difficulty with the continuous flavor symmetries described above is that
they contain few low dimensional representations, and therefore there are few
invariant tensors that are of use in building up the fermion mass matrix in
powers of $\epsilon$.  In contrast, if one is willing to consider nonabelian
discrete groups for $G_f$ one can find groups with an arbitrarily large number
of triplet representations, for example. With such a symmetry there are many
invariant tensors which can arise without resorting to a multitude of exotic
particles.  In this paper we consider the $\Delta(3n^2)$ dihedral subgroups of
$SU(3)$, which contain an arbitrary number of triplet representations.  The
explicit model we give is based on  $\Delta(75)$, a group with eight triplet
and three singlet representations.

\newsec{Nonabelian discrete symmetries}

The representations of discrete groups with ${}^\circ G$ elements satisfy the
relation $\sum_i d_i^2 =  {}^\circ G$, where $d_i$ is the dimension of the
$i^{th}$ representation.  Thus finite groups have a finite number of finite
dimensional representations.  Among the nonabelian discrete groups most
familiar to physicists, namely the crystallographic symmetries,  the ones with
more than one triplet representation are the octahedral, and icosahedral
groups.  The octahedral group $O$ has 24 elements and representations
$\{1,1',2,3,3'\}$.  We could consider constructing an $SU(5)\times O$ grand
unified theory, for example, by having the $Q$, $U$ and $E^c$  fermions
transform as a (10,3).  However one finds that
\eqn\octa{3\otimes 3 = 3_a\oplus 3'_s \oplus 2_s \oplus 1_s\ .}
Evidently the 5 of $SO(3)$ decomposes as a $3'_s \oplus 2_s$ under $O$.  This
does not help to solve the problem encountered with $SO(3)$ as a flavor group,
since each of these couplings leads to a rank two mass matrix again:  the $3'$
and 2 decompositions of $3\otimes 3$ consist of
\eqn\octdec{(3\otimes 3)\vert_{3'} = \(\matrix{3 \lambda_6 3\cr
3 \lambda_4 3\cr 3 \lambda_1 3\cr}\)\qquad    (3\otimes 3)\vert_{2} =
\(\matrix{ 3 \lambda_3 3\cr 3 \lambda_8 3\cr}\)\ ,}
where the $\lambda_a$ are the Gell-Mann $SU(3)$ matrices and $3 \lambda_a 3 =
3_i (\lambda_a)_{ij} 3_j$.  The same conclusion holds for the icosahedral
group.

What is needed to explain the top mass operator \upmss\ is a group which
contains a triplet $3=\{x,y,z\}$ as well as a $3'$ representation contained in
$3\otimes 3$ with $3\otimes 3 \vert_{3'}=\{x^2,y^2,z^2\}$.  Then the top mass
arises at tree level  if the Higgs transforms as  $H_u={3'}^*$ with a vev only
in the third family component.  This is only possible if the $3$ representation
is complex, since otherwise $x^2 + y^2 + z^2$ is a singlet.  It follows that
$G_f$ cannot be a subgroup of $SO(3)$, and we turn to discrete subgroups of
$SU(3)$\foot{All of our discussion of discrete $SU(3)$ subgroups is based on
ref. \sut.}.

The discrete subgroups of $SU(3)$ are the irregular groups $\Sigma$ and the
dihedral groups $\Delta(3n^2)$ and $\Delta(6n^2)$ for all integers $n$.  The
$\Delta(3n^2)$ groups are particularly interesting since their representations
consist solely of triplets and singlets. These groups are of order $3n^2$ and
are generated by the matrices
\eqn\egen{E_{00} = \[\matrix{0&1&0 \cr 0&0 &1\cr 1&0&0\cr}\]}
and
\eqn\agen{A_{pq} = \[\matrix{(\eta_n)^p&0 &0 \cr 0& (\eta_n)^q&0 \cr 0&0&
(\eta_n)^{-(p+q)}}\]\ ,}
where $\eta_n$ is the $n^{th}$ root of unity
\eqn\nroot{\eta_n = {\rm e}^{2\pi i/n}\ ,}
and $p,q$ are integers.

The irreducible representations of the $\Delta(3n^2)$ groups consist of
(i)  9 singlets and  $(n^2-3)/3$ triplets for $n$ a multiple of three;
(ii) 3 singlets and $(n^2-1)/3$ triplets otherwise.
The large number of inequivalent triplet representations in these groups are
invaluable for building a  model of fermion masses, starting with flavor
democracy at short distances. In this paper we will focus on a particular
discrete symmetry in order to exhibit some of the general features of model
building with nonabelian discrete symmetries.  The symmetry we discuss is
$\Delta(75)$ (\ie, $\Delta(3n^2)$ with $n=5$), which is apparently the smallest
of the dihedral groups with sufficient structure to be interesting.

\subsec{$\Delta(75)$}

The irreducible representations of $\Delta(75)$  include one real singlet
$A_1$, one complex singlet  $A_2$, and  four complex triplets $T_1\ldots,T_4$.
 The  character table may be constructed from the generators \egen, \agen\ with
$n=5$ and is given in Table 1. (For an explanation of discrete symmetries and
character tables see, for example, ref. \ref\dgt{D.B. Chesnut, {\it Finite
Groups and Quantum Theory}, Robert E. Kreiger Co., Malabar Florida, 1982;
M. Hamermesh, {\it Group Theory and its Application to Physical Problems},
Dover Pub. Inc., New York, 1989}.)
\topinsert
\def\spacer{height2pt&\omit&&
\omit&\omit&\omit&\omit&\omit&\omit&\omit&
\omit&\omit&\omit&\omit&\omit&\omit&\omit&
\omit&\omit&\omit&\omit&\omit&\omit&\omit&}
\vbox{
\offinterlineskip
\hrule
\halign {
\vrule#&\strut \quad\hfil # \quad
&&\vrule# &\strut\hskip \gapi \hfil #\hfil\hskip \gapi \cr
\spacer\cr
&$\Delta(75)$&& $E$&\omit
&$3A_{10}$&\omit&3$A_{20}$&\omit&3$A_{30}$&\omit&3$A_{40}$&\omit&
3$A_{11}$&\omit&3$A_{22}$&\omit&3$A_{33}$&\omit&3$A_{44}$&\omit&
25C&\omit&25E&\cr
\spacer\cr
\noalign{\hrule}\spacer\cr
&$A_1$ && 1 &\omit& 1 &\omit&  1 &\omit& 1 &\omit& 1 &\omit& 1
&\omit& 1 &\omit& 1 &\omit& 1 &\omit& 1 &\omit& 1 & \cr

&$A_2$ && 1 &\omit& 1 &\omit&  1 &\omit& 1 &\omit& 1 &\omit& 1
&\omit& 1 &\omit& 1 &\omit& 1 &\omit& $\omega$ &\omit& $\bar\omega$ & \cr

&$T_1$ && 3 &\omit&
$\chi_{_{10 }}$ &\omit& $\chi_{_{20 }}$ &\omit&  $\chi_{_{20 }}$ &\omit&
$\chi_{_{10 }}$ &\omit& $\chi_{_{11 }}$ &\omit&  $\chi_{_{22 }}$ &\omit&
$\bar\chi_{_{22 }}$ &\omit& $\bar\chi_{_{11 }}$
&\omit& 0 &\omit& 0 & \cr

&$T_2$ && 3 &\omit&
$\chi_{_{20 }}$ &\omit& $\chi_{_{10 }}$ &\omit&  $\chi_{_{10 }}$ &\omit&
$\chi_{_{20 }}$ &\omit& $\chi_{_{22 }}$ &\omit&  $\bar\chi_{_{11 }}$ &\omit&
$\chi_{_{11 }}$ &\omit& $\bar\chi_{_{22 }}$
&\omit& 0 &\omit& 0 & \cr

&$T_3$ && 3 &\omit&
$\chi_{_{11 }}$ &\omit& $\chi_{_{22 }}$ &\omit&  $\bar\chi_{_{22 }}$ &\omit&
$\bar\chi_{_{11 }}$ &\omit& $\chi_{_{20 }}$ &\omit&  $\chi_{_{10 }}$ &\omit&
$\chi_{_{10 }}$ &\omit& $\chi_{_{20 }}$
&\omit& 0 &\omit& 0 & \cr

&$T_4$ && 3 &\omit&
$\chi_{_{22 }}$ &\omit& $\bar\chi_{_{11 }}$ &\omit&  $\chi_{_{11 }}$ &\omit&
$\bar\chi_{_{22 }}$ &\omit& $\chi_{_{10 }}$ &\omit&  $\chi_{_{20 }}$ &\omit&
$\chi_{_{20 }}$ &\omit& $\chi_{_{10 }}$
&\omit& 0 &\omit& 0 & \cr
\spacer\cr}
 \hrule}
\caption{{\bf Table 1.} {\it Character table for $\Delta(75)$, computed from
ref. \sut. The quantities $\chi$ and $\omega$ are defined as $\chi_{pq}=
(\eta_5)^p + (\eta_5)^q +(\eta_5)^{-p-q}$, and $\omega= \eta_3$, where
$\eta_n={\rm e}^{2\pi i/n}$. }}

\endinsert
The defining representation is taken to be $T_1$, and we have labelled the
conjugacy classes after generators contained in that class for the $T_1$
representation. For example, the class labelled $3A_{10}$ contains the group
elements
$A_{10} $, $ A_{04}$, and $A_{41}$
\eqn\classi{
\[\matrix{(\eta_5)^1&0 &0 \cr 0& (\eta_5)^0&0 \cr 0&0& (\eta_5)^{4}}\] ,\ \
\[\matrix{(\eta_5)^0&0 &0 \cr 0& (\eta_5)^4&0 \cr 0&0& (\eta_5)^{1}}\]  ,\ \
\[\matrix{(\eta_5)^4&0 &0 \cr 0& (\eta_5)^1&0 \cr 0&0& (\eta_5)^{0}}\] ,}
 in the $T_1$ representation, while the class $25E$ contains the 25 elements
\eqn\classii{E_{pq}  = \[\matrix{0&\eta_5^p&0 \cr 0&0 &\eta_5^q\cr
\eta_5^{-(p+q)}&0&0\cr}\]\ .}
The  $25C$ class contains  the square of the $E_{pq}$ matrices.

{}From  the character table it is possible to determine the decomposition of
the product of any two representations.  Evidently $A_1$ is the trivial
representation, while
\eqn\sing{ A_2\otimes A_2 = \bar A_2\ ,\qquad A_2\otimes \bar A_2 = A_1\
,\qquad A_2 \otimes T_i = \bar A_2\otimes T_i = T_i\ ,}
where $i=1,\ldots,4$.  Less obvious are the products of two triplet
representations, whose decompositions are given in Table 2.
\topinsert
\def\spacerii{height3pt&\omit&&\omit&&
\omit&\omit&\omit&\omit&\omit&\omit&\omit&
\omit&\omit&\omit&\omit&\omit&\omit&\omit&\omit&}
\def\spaceriii{height1pt&\omit&&\omit&&
\omit&\omit&\omit&\omit&\omit&\omit&\omit&
\omit&\omit&\omit&\omit&\omit&\omit&\omit&\omit&}

\vbox{
\offinterlineskip
\hrule
\halign{
\vrule#&\strut \quad\hfil # \hfil\quad &\vrule# &\strut \hskip1pt #
&&\vrule# &\strut \hskip\gapii\hfil #\hfil\hskip\gapii\cr
\spacerii\cr
&$\Delta(75)$&&&& $1$&\omit
&$ \bar 1 $&\omit&$ 2 $&\omit&$ \bar 2 $&\omit&$ 3 $&\omit&
$ \bar 3 $&\omit&$ 4 $&\omit&$\bar 4 $&\cr
\spacerii\cr
\noalign{\hrule}\spaceriii\cr
\noalign{\hrule}\spacerii\cr
&$1$&&&& $\bar 1 \bar 1 2$&\omit
&$A3\bar 3$&\omit&$\bar 2 3 \bar 3$&\omit&$\bar 1 4 \bar 4$
&\omit&$1 \bar 2 4$&\omit&
$1 \bar 2 \bar 4$&\omit&$2 3 4$&\omit&$2 \bar 3 \bar 4$&\cr

&$2$&&&&$\bar 2 3 \bar 3$&\omit
&$1 4 \bar 4$&\omit&$\bar 1 \bar 2\bar 2$&\omit&$A 4 \bar 4$
&\omit&$\bar 1 3 \bar 4$&\omit&
$\bar 1 \bar 3 4$&\omit&$1 2 \bar 3 $&\omit&$1 2 3$&\cr

&$3$&&&&$1 \bar 2 4$&\omit
&$\bar 1 2  4$&\omit&$\bar 1 3 \bar 4$&\omit&$1 3 \bar 4$
&\omit&$\bar 3 \bar 3 4$&\omit&
$A 2 \bar 2 $&\omit&$ 2 \bar 2  \bar 4 $&\omit&$1 \bar 1 \bar 3$&\cr

&$4$&&&& $2 3 4$&\omit
&$\bar  2 3 4$&\omit&$1 2 \bar 3 $ &\omit&$\bar 1\bar 2\bar 3$
&\omit& $ 2 \bar 2  \bar 4 $&\omit&
$1 \bar 1   3$ &\omit&$ 3 \bar 4  \bar 4 $&\omit&$A 1 \bar 1 $&\cr
\spacerii\cr}
 \hrule}

\caption{{\bf Table 2.} {\it Decomposition of the product of two triplets.
Triplets $T_n$ and $\bar T_n$ are represented by $n$ and $\bar n$ respectively,
while $A\equiv A_1 \oplus A_2\oplus \bar A_2$.  For example, $T_3\otimes \bar
T_1 = \bar T_1 \oplus T_2 \oplus T_4$, and  $T_1\otimes \bar T_1 = A_1 \oplus
A_2\oplus \bar A_2 \oplus T_3 \oplus \bar T_3$.}}
\endinsert

Since we wish to construct explicit models of particle couplings obeying
$\Delta(75)$ symmetry, we need to choose a basis for all of the representations
and  construct the invariant tensors.  We have chosen a basis defined by
\eqn\basdef{T_1\otimes T_1\vert_{T_2} = \trip{x^2} {y^2}{ z^2} \ ,\quad
T_1\otimes\bar T_1\vert_{T_3} = \trip{ y\bar z}{ z\bar x}{ x\bar y}\ ,\quad
T_2\otimes\bar T_2\vert_{T_4} = \trip{ b\bar c}{ c\bar a}{ a\bar b}\ ,}
where we have written $T_1 = \{x,y,z\}$, $T_2 = \{a,b,c\}$.  This basis has the
virtue that the generator $E_{00}$ is the same matrix \egen\ in all of the
triplet representations.  Thus when any two triplets $T_i$ and $T_j$ (or their
conjugates) are combined into a third triplet $T_k$, the elements of $T_k$ must
cyclically permute when the elements of $T_i$ and $T_j$ are simultaneously
cyclically permuted; therefore all of the components of $T_k$ are specified
when the first component is known.  The decomposition of all products of
triplets in this basis are given in the appendix.

\subsec{Symmetry breaking}

We now turn to ways to spontaneously break the $\Delta(75)$ symmetry in a
supersymmetric theory.  One reason we choose to focus on supersymmetry is that
the flavor breaking patterns can be more interesting:  in a supersymmetric
theory one can have different symmetry breaking patterns in different sectors
of the theory which communicate only through higher dimension operators and not
through radiative corrections.  Non-generic flavor symmetry
breaking  can lead to interesting structure, as we will show.  Here we give a
couple of toy models showing different symmetry breaking patterns.

The first toy model we consider has $\Delta(75)$ breaking down to $Z_3$
generated by $E_{00}$ alone (eq. \egen). We include the singlet fields $S$,
$\phi$, $\bar \phi$ transforming as the $A_1$, $A_2$ and $\bar A_2$
representations respectively, as well as $Z$ and $\bar Z$ triplets transforming
as $T_1$ and $\bar T_1$.  The (nonrenormalizable) superpotential is taken to be
\eqn\toywi{W=\alpha S (-3\mu^2 + \bar Z Z) + \beta \phi \bar Z Z + \gamma
\bar\phi\bar Z Z + {g\over 3}\bar Z^3 + { Z^5 \over 5 M^2}\ .}
Written in terms of  components, the above interactions read (see the appendix)
\eqn\toywic{\eqalign{
W&= \alpha S (-\mu^2 + \bar Z_1Z_1+ \bar Z_2 Z_2 +\bar Z_3 Z_3) +
\beta  \phi(\bar Z_1Z_1+ \omega \bar Z_2 Z_2 +\omega^2 \bar Z_3 Z_3)\cr
 &+
\gamma \bar \phi(\bar Z_1 Z_1+ \omega^2 \bar Z_2 Z_2 +\omega  \bar Z_3 Z_3) +
g(\bar Z_1\bar Z_2 \bar Z_3 ) \cr
&+ (Z_1^5 + Z_2^5 + Z_3^5)/5 M^2}}
(where $\omega \equiv e^{2i\pi/3}$)
with several isolated supersymmetric minima; all have $\phi = \bar\phi =0$. One
of the vacua takes the values
\eqn\toyimin{Z=\mu\delta \trip{1}{1}{1}\ ,\quad \bar Z = {\mu\over 3
\delta}\trip {1}{1}{1}\ ,\quad S=-{g\mu\over 9\alpha\delta^3}\ ,}
with
$$\delta=\[ g M^2\over 27 \mu^2\]^{1/8}\ .$$

Our second example has $\Delta(75)$  broken to $Z_5$ by giving a triplet a vev
in a single component.  The toy model includes the following superfields that
transform as irreducible representations under $\Delta(75)\times U(1)$, where
the $U(1)$ is gauged:
\eqn\toywii{S= (A_1)_0\ ,\quad Z= (T_1)_1\ ,\quad \bar Z = (\bar T_1)_{-1}\
,\quad R= (T_1)_{-2}\ ,\quad \bar R= (\bar  T_1)_{2}\ .}
{}From these fields we construct the renormalizable superpotential
\eqn\toycom{
 W = \alpha S (-\mu^2 + \bar Z Z) - M R \bar R  + \beta   R Z Z  +\gamma\bar
R\, \bar Z\, \bar Z\ .}
 In terms of component fields,
\eqn\toywiic{\eqalign{
W&=\alpha S (-\mu^2 + \bar Z_1 Z_1 +c.p.)
 -M(  \bar R_1 R_1  + c.p. )
  \cr &+\beta  (R_1  Z_2  Z_3  +c.p. ) +
\gamma  (\bar R_1  \bar Z_2  \bar Z_3  + c.p.)\ , \cr}}
where {\it c.p.} stands for cyclic permutation of each triplet's indices (see
appendix).
Minimizing the scalar potential (including the $D$-term from the gauged
$U(1)$)yields three families of supersymmetric vacua, including the isolated
solution
\eqn\vacii{
S=R=\bar R=0,\qquad Z = \bar Z = \trip{0}{0}{\mu}\ . }

\subsec{Fermion mass texture}

Flavor symmetry breaking can be communicated to the Yukawa couplings of the
light fermions in two ways: either through the mixing of light and heavy
fermions, or through the Higgs potential. We have seen that in flavor
unification, the large top quark mass requires that the Higgs fields $H_u$
transform under flavor at short distances and have direct (unsuppressed) flavor
symmetry breaking vevs.  Keeping in mind that the successful GUT prediction for
$\sin^2\theta_w$ assumes that there are only two Higgs doublets below the GUT
scale, it is natural to suppose that flavor symmetry breaking occurs at the GUT
scale or above, and that all but these two Higgs doublets acquire large masses.

For example,
suppose $H_u$ and $H_d$ are Higgs doublets that are both  flavor triplets in
the $\bar T_2$ and $\bar T_1$ representations of $\Delta(75)$ respectively, and
that they couple to the left-chiral superfield triplets  $Z=T_3$ and $\bar
Z=\bar T_3$, which are gauge singlets.  There are two couplings,
\eqn\muterm{\eqalign{
W &= \lambda Z H_u H_d + \lambda' \bar Z H_u H_d \cr
    &= \lambda (Z_1  H_{u2} H_{d1} + c.p.) + \lambda' (\bar Z_1  H_{u3} H_{d1}
+ c.p.) \ .}}
If $Z$ and $\bar Z$ get the vevs $\{\mu,0,0\}$ and $\{0,\mu,0\}$  respectively,
where $\mu$ is some very heavy scale, then only the Higgs doublets $H_{u3}$ and
$H_{d3}$ remain light and are able to eventually develop $SU(2)\times U(1)$
breaking vevs. What has happened is that $\Delta(75)\times U(1)_{PQ}$ has been
broken down to a diagonal $Z_5$, where $U(1)_{PQ}$ is the Peccei-Quinn symmetry
in the interactions \muterm.  The three components of both of the Higgs
doublets carry $Z_5$ charges
that allow two of the Higgs flavors to pair up and become heavy, while
protecting the third.

We now incorporate these ideas into a toy model based on $\Delta(75)\times
U(1)$ that leads to an interesting fermion mass hierarchy, ignoring gauge
interactions for the moment. The ``matter'' fields are
$$ F= (T_1)_1\ , \quad \psi = (\bar T_4)_1\ ,\quad \bar \psi = ( T_4)_1 $$
where $F$ will play the role of three families of quarks and leptons, while
$\psi$ and $\bar \psi$ are three vectorlike exotic families that will become
heavy when the $U(1)$ is broken.  This occurs at a scale $M$ when the singlet
field $S$ develops a vev:
$$ S = (A_1)_{-2}  =  M\ . $$
At a somewhat lower scale  $\Delta(75)$ is broken, and we assume that this is
due to the fields
$$X=(\bar T_3)_{-2} = x M \trip{1}{1}{1}\ ,\quad Y=(\bar T_1)_{-2} = y M
\trip{1}{1}{1}\ ,$$
where $x$ and $y$ are small numbers.
The fermions $F$ only get a mass when the ``Higgs'' field $H$ gets a vev, and
we assume that
$$H = (\bar T_2)_{-2} = \trip {0}{0}{v}\ ,$$
where $v \ll M$ is the ``weak scale'',  envisaging a mechanism such as
described above that renders all but the third family component of $H$ heavy at
the scale $M$.

The most general renormalizable superpotential $W_m$ describing the
interactions of the matter fields with $S$, $X$, $Y$, and $H$ is given by
\eqn\toywiii{\eqalign{
W_m &= S\bar\psi \psi + X\bar\psi F + Y\bar\psi\psi + H(FF + F \psi)\cr
&= S(\bar\psi_1 \psi_1) + X_1\bar\psi_3 F_3 + Y_1\bar\psi_3\psi_2 + H_3(F_3F_3
+ F_2 \psi_3) + c.p.
\ .}}
(For simplicity we have omitted coupling constants, assumed to all be
$\CO(1)$).
At the scale $M$ the $\psi$ field gets a mass and is integrated out of the
theory, giving rise to the effective theory
\eqn\toyeff{W_{eff} = Y_{ij} H_3 F_i F_j\ .}
\topinsert
\centerline{\epsfbox{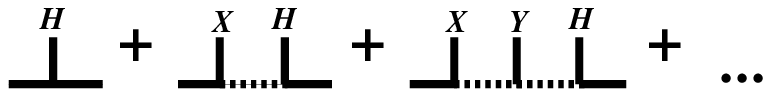}}
\caption{{\bf Fig. 1.} {\it Leading  supergraph contributions to the effective
Yukawa coupling of the $F$ superfield in eq. \toyeff .  The internal dotted
lines indicate $\psi$ and $\bar \psi$ superfields with mass M.  The unlabelled
external lines are the light fermions $\CF$.}}
\endinsert
The Yukawa coupling $Y_{ij}$ can computed by summing the diagrams in fig. 1,
making use of the invariant tensors discussed in the appendix. The result is
\eqn\toyuk{ Y_{ij} \sim \[\matrix{0 & x y^2 & 0\cr x y^2 & xy & x\cr 0 & x &
1\cr}\]\ .}
In addition there are wavefunction renormalization graphs which give effective
$D$-terms which eliminate the zeros in the above matrix, but they are
negligible:  the $\{13\}$ and $\{31\}$ entries in $Y_{ij}$ receive $\CO(\vert
x\vert^2 y^*)$ contributions, while the $\{11\}$ entry is $\CO(\vert x\vert^4
y^{*2})$.
 $Y_{ij}$ exhibits an obvious hierarchical structure, and with $x\sim y\sim
1/20 $, it could provide a reasonable description of the Yukawa coupling matrix
of the up-type quarks at the GUT scale \bbo.    In the next section we
incorporate this toy model into  $SO(10)$ and $SU(5)$ grand unified theories.

\pag
\newsec{A supersymmetric $SO(10)\times \Delta(75)$ GUT}

In this section we show how to use nonabelian discrete flavor symmetries to
construct a GUT in which the gauge and flavor symmetries are separately
unified.  In particular, we show how to incorporate the the toy model \toywiii\
into an $SO(10)$ grand unified theory.  To get realistic quark masses it is
necessary that the $Y_D$ Yukawa coupling of the down quark matrix look quite
different from $Y_U$; we achieve this by having the Higgs fields $H_u$ and
$H_d$ transform as different flavor representations.  The representations are
chosen so that (i) down-type quarks get masses at higher order in symmetry
breaking, explaining the small $b/t$ mass ratio  without requiring unnaturally
large $\tan\beta$; (ii) the $\{22\}$ and $\{12\}$ entries of the down mass
matrix are susceptible to large corrections from higher dimension operators
which arise from Planck scale physics, accounting for  $m_s/m_b \gg m_c/m_t$
and the large Cabbibo angle.

\subsec{Fields and interactions}
The model we offer as an example is an $SO(10)\times \Delta(75)$
supersymmetric GUT, where $\Delta(75)$ is the flavor group \foot{$SO(10)$ GUTS
have been discussed extensively in the literature.  See \ref\hrr{J.A. Harvey,
D.B. Reiss and P. Ramond, \np{199}{1982}{223}}, and for recent references,
\ref\soten{ G. Anderson {\it et al.}, preprint LBL-33531, hep-ph-9308333; R.N.
Mohapatra, preprint hep-ph/9310265}.}.  This example is an extension of the
toy model \toywiii, containing both ``matter superfields'' which do not get
vevs, and ``Higgs superfields'' which do.  The matter fields consist of three
ordinary chiral families
\eqn\ord{ \CF  =(16,T_1)\ ,}
as well as exotic fields:
\eqn\exotic{ \psi   = (16, \bar T_4)\ ,\quad \bar \psi   = (\bar{16},  T_4)\
,\qquad
\chi=   (10,\bar T_2)\ ,\quad \bar{\chi} = (10,T_2)\ .}

There are several fields associated with symmetry breaking. To break $SO(10)$
down to  $SU(3)\times SU(2)\times U(1)$ at $M_{GUT}\simeq 10^{16}$ GeV  in the
most economical fashion
requires both a 45 and a 16  of Higgs, and we include a conjugate partner for
the latter.  These fields are assumed to come in $\Delta(75)$ triplets:
\eqn\sohiggs{\Sigma=(45,\bar T_4)\ ,\quad \Omega = (16, T_2)\ ,\quad \bar
\Omega = (\bar {16},\bar T_2)\ .}
There are also gauge singlets which get vevs at a similar scale, namely
\eqn\gsing{X=(1,\bar T_3)\ ,\qquad Y= (1,\bar T_1)\ ,\qquad Z=(1,T_2)\ .}
Finally there are singlet fields $S$ and $S'$ which are invariant under both
$SO(10)$ and $\Delta(75)$; their vevs are responsible for the masses of the
vectorlike fermion families $\psi$ and $\chi$, and occur over an order of
magnitude above $M_{GUT}$.

To break the weak interactions we require a $10$ of Higgs; we will take three
families of these Higgs as well. In order to construct a model without the
fine-tuning problems associated with large $\tan \beta = \vev{H_u/H_d}$
\ref\tanbeta{T. Banks, \np{303}{1988}{172}; L.J. Hall , R. Rattazzi and U.
Sarid, LBL-33997 (1993), hep-ph-9306309;
 A.E. Nelson and L. Randall, UCSD-PTH-93-24/MIT-CTP-2230 (1993),
hep-ph-9308277.}, we have the up and down Higgs doublets reside in different
10's:
\eqn\whiggs{ H_u = (10,\bar T_2)\ ,\qquad H_d = (10,\bar T_1)\ .}
As we will show below, the flavor quantum numbers of $H_d$ are chosen so that
the down type quarks have naturally suppressed Yukawa couplings.

$SO(10)\times \Delta(75)$ symmetry allows us to write down the renormalizable
superpotential
\eqn\wpot{\eqalign{
 W_m &= S \bar \psi   \psi   +  S'\bar\chi\chi  \cr
&+ X\bar \psi   \CF   + Y\bar \psi   \psi   + \chi\[\CF   \CF   + \CF   \psi
\]\cr
&+H_u \[ \CF   \CF   + \CF    \psi   \]
+ H_d \bar\chi Y  \ .}}
For notational simplicity we have not indicated coupling constants for these
operators, which are all assumed to be $\CO(1)$.  Note that we have omitted a $
S\bar\chi H_u$ operator, which can be done by choosing suitable definitions of
the $\chi$ and $H_u$ fields, which have the same quantum numbers. Other
operators allowed by $SO(10)\times \Delta(75)$ but absent from \wpot, such as
$M_p \bar \psi   \psi $, operators involving $Z$, $\Sigma$ and $\Omega$, etc,
may be naturally excluded by imposing an additional $U(1)$ or $Z_N$ symmetry to
the theory which commutes with flavor and has no $SO(10)$ anomalies.  The
choices of charges under this symmetry are not unique, and in fact the symmetry
can be either an $R$-symmetry or ordinary.  It is the spontaneous violation of
this abelian symmetry by $\vev{S}$ and $\vev{S'}$ that determines the masses of
the heavy fermions $\psi$ and $\chi$.

Although the fields $\Sigma$, $Z$ and $\Omega$ do not have renormalizable
couplings to the matter fields $\CF$, $\psi$ and $\chi$,   they will interact
through operators of dimension five and higher suppressed by powers of  $ M_p$.
 By means of the same Abelian symmetry controlling operators in the
renormalizable sector of the theory, the allowed dimension five operators can
be restricted to
\eqn\dimfive{
W_{grav.} = {1\over M_p}\[ H_d \CF\CF Z+ \Sigma H_d \CF\CF  +\CF\CF\bar\Omega\
\bar\Omega \]\ .}
As we will show below, the first two operators give important contributions to
the down quark mass matrix, while the third operator is responsible for giving
an interesting pattern of  neutrino masses.  Furthermore, in an $SU(5)$ version
of this model, the second operator can explain the ratio of down quark masses
to charged lepton masses {\it \`a la} Georgi-Jarlskog \ref\gj{H. Georgi and C.
Jarlskog,
\pl{86}{1979}{297}
}.

In order to generate realistic masses for the quarks and leptons, it is
necessary to make certain assumptions about the symmtery breaking pattern of
the fields that get vevs.  We make the following assumptions, along the lines
of our discussion of symmetry breaking in the previous section:
\item{1.} The $S$ and $S'$ fields get vevs at a scale which is about $20 - 50
M_{GUT}$, giving large masses to the $\psi$ and $\chi$ fields.
\item{2.} The $X$, $Y$ and $Z$ fields get vevs on the order of $M_{GUT}$ in
each component, inducing mass mixing between the heavy fermions $\psi$, $\chi$
and the light fermions $F$.
\item{3.} $SO(10)$ is broken to $SU(3)\times SU(2)\times U(1)$ at the GUT scale
by vevs of the $\Omega$, $\bar \Omega$ and $\Sigma$ fields.  We assume that
each flavor component of the $\bar\Omega$ and at least  the second flavor
component of $\Sigma$ develop vevs.
\item{4.}  Of the $H_u$ and $H_d$ triplets, only the $Y=-1/2$ weak doublet from
$(H_u)_3$ and the $Y=+1/2$ weak doublet from $(H_d)_3$
remain lighter than $M_{GUT}$ and develop $SU(2)\times U(1)$ breaking vevs.

The reason we take the flavor symmetry breaking scale to be so high is dictated
by the desire to keep interactions perturbative up to scales near the Planck
mass.  This is a generic feature of models of flavor unification where masses
arise through mixing with heavy fermions:  such theories will have at least an
extra set of fermion families as well as their mirrors which, with the Higgs
fields,  render the gauge theory asymptotically unfree above the flavor
unification scale.  Thus the scale of flavor physics is forced to lie within a
few decades of the Planck scale.  Furthermore, it is interesting to note that
gauge interactions are often strong very near the scale where quantum gravity
is expected to be relevant.

\subsec{Quark masses}

The effective quark  Yukawa couplings are generated in this model when the
$\psi$ and $\chi$ fields are integrated out of the theory at the scales
$\vev{S}$ and $\vev{S'} $ --- taken to lie above $M_{GUT}$ --- and the symmetry
breking fields $X$, $Y$, $Z$, and $\Sigma$ acquire their vevs.  The diagrams
arising from the renormalizable interactions \wpot\ that contribute to an
effective superpotential are shown  in Fig. 2.
\topinsert
\centerline{\epsfbox{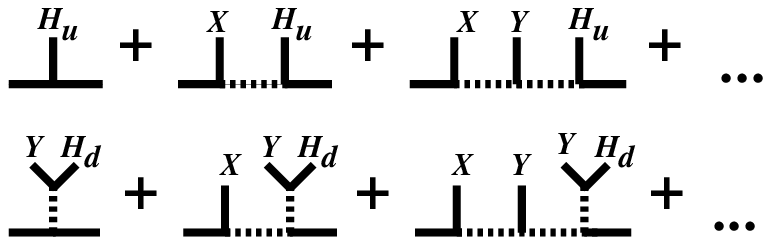}}
\caption{{\bf Fig. 2.} {\it Leading  supergraph contributions to quark and
lepton Yukawa couplings.  The internal lines indicate $\psi$, $\bar \psi$,
$\chi$ and $\bar \chi$ superfields.  The unlabelled external lines are the
light fermions $\CF$.}}
\endinsert
Denoting
$$\vev{X/S} \equiv x\ ,\qquad \vev{Y/S}\equiv y\ ,\quad \vev{Y/S'} \equiv y'$$
 and ignoring both the $\CO(1)$ coefficients in \wpot, the effective Yukawa
couplings  generated from these diagrams are
\eqn\yeffi{
Y_u \sim \(\matrix{
{0} &{xy^2} & {0}\cr
{xy^2}& {xy}& {x}\cr
{0}& {x}& {1}\cr}\) \ ,\qquad
Y_d \sim  y'\(\matrix{
{0} &{xy^2} & {0}\cr
{xy^2}& {xy}& {x}\cr
{0}& {x}& {1}\cr}\) \ ,}
where $Y_u$ and $Y_d$ are the coefficients of the effective operators
$H_u\CF\CF $ and  $H_d\CF\CF $ respectively.  One sees that there is a natural
hierarchical structure to the masses, and that down-type quarks are
automatically a factor of $y'$ more weakly coupled to the Higgs doublet than
are up-type quarks.  The two matrices are not simply proportional to each other
(due to the omitted $\CO(1)$ coefficients of \wpot), so that there are nonzero
mixing angles, although there may be partial cancellations leading to a small
$\theta_{23}$.

Additional important contributions to $Y_u$ and $Y_d$ come from the dimension
five operators \dimfive, which enter the effective Yukawa couplings through the
diagrams pictured in Fig. 3.
\topinsert
\centerline{\epsfbox{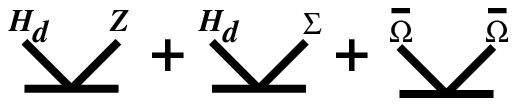}}
\caption{{\bf Fig. 3.} {\it Supergraphs involving the dimension five operators
\dimfive\ contributing to quark and lepton Yukawa couplings.
  }}
\endinsert
The  first two graphs in fig. 3 contribute to the  $d$ and $s$ quark masses, as
well as the Cabbibo angle.
Denoting
$$ \delta_z \equiv \vev{Z/\Lambda}\ , \qquad \delta_{\Sigma} \equiv
\vev{\Sigma/ M_p}\ ,$$
eq. \yeffi\ is modified to read
\eqn\yeffi{
Y_u \sim \(\matrix{
0 &{xy^2} & {0}\cr
{xy^2}& {xy}& {x}\cr
{0}& {x}& {1}\cr}\) \ ,\qquad
Y_d \sim  \(\matrix{
0 &{\delta_z} & 0\cr
{\delta_z}& { \delta_{\Sigma}}& {xy'}\cr
0& {xy'}& {y'}\cr}\) \ }
for the Yukawa couplings at the GUT scale.  We have only given the leading
contributions to each entry, and ignore the negligible contributions from
wavefunction renormalization to the $\{13\}$, $\{31\}$ and $\{11\}$ entries.
Taking scaling effects into account,
these matrices can lead to realistic quark masses for the values
$$x\sim y\sim \frac{1}{20}\ ,\quad y'\sim \frac{1}{50}\ $$
and imply
$$\tan\beta\simeq 3\ ,$$
for a top quark mass $m_t \simeq 160$ GeV.  This fit assumes that the couplings
in $W_m$ \wpot\ are all $\CO(1)$  and works best if the   couplings in
$W_{grav.}$ \dimfive\ are actually $\simeq 0.5$ (\ie, so that the
characteristic scale of nonrenormalizable gravitational interactions is
$2M_p$.).

\subsec{Lepton masses}
The third diagram in fig. 3 gives the right-handed neutrino a Majorana mass
\eqn\majmass{ M_{\nu} \sim  { \vev{\bar \Omega}^2 \over  M_p }
\times
\( \matrix{0 &1&1\cr 1&0&1\cr 1&1&0\cr} \)\ , }
where the entries denoted as ``1'' are to be understood as $\CO(1)$.  By
identifying the $B-L$ breaking scale with the GUT scale, the fact that $\CF$
couples to $\bar \Omega$ only through a dimension five operator naturally
predicts a Majorana mass of $M_{GUT}^2/ M_p$.  The seesaw mechanism
\ref\seesaw{M. Gell-Mann, P. Ramond, R. Slansky  in {\it Supergravity} (ed.
D.Z. Freedman and P. van Nieuwenhuizen)
North Holland, Amsterdam}\
then leads to a tau neutrino mass of roughly $ M_p/(G_F M_{GUT}^2)$ --- where
$G_F$ is the Fermi constant --- which gives rise to a mass hierarchy for
neutrinos that is of interest both for dark matter and neutrino oscillations.

The charged lepton masses do not work in the $SO(10)$ model described above,
but do in a similar $SU(5)$ version, where $\CF\to \bar 5 + 10 + 1$, $H_d\to
\bar 5$, $H_u\to 5$, $\Sigma\to 24$ and so forth.   In this model the $\{22\}$
entry in $Y_d$ in eq. \yeffi\  involves $SU(5)$ breaking through the coupling
to the $\bar 5 \oplus \bar {45}$ in $H_d \Sigma = \bar 5 \times 24 $.   If the
coupling is primarily in the $\bar{45}$ channel, then the mass matrices are
similar to the Georgi-Jarlskog form and yield the successful  GUT-scale mass
relations \gj
\eqn\gjrel{{m_b\over m_\tau} \simeq  1\ ,\qquad {m_s\over m_\mu} \simeq
{1\over 3}\ ,\qquad {m_d\over m_e} \simeq 3\ .}
 We do not bother writing down the $SU(5)$ model, since it is in almost every
respect identical to the $SO(10)$ version described above.
The reason why the Georgi-Jarlskog mechanism doesn't work in the $SO(10)$
version of the model is that $H_d\Sigma=10\times 45$ can only couple to
$\CF\CF$ as a $10$, which does not split down quark from lepton masses.

\newsec{Flavor changing neutral currents}

In the standard model flavor changing neutral currents (FCNC) must proceed
through dimension six operators, and so experiments are insensitive to physics
above $\sim 1000$ TeV.  In contrast, FCNC enter supersymmetry through dimension
two squark mass matrices, and are sensitive to physics at very short distances
\ref\sfcnc{L.J. Hall, V.A.Kostelecky and S. Raby, \np{267}{1986}{415}; H.
Georgi, \pl{169}{1986}{231}}.  Limits on FCNC from the neutral $K$ and $B$
mesons require that the squarks must be mass eigenstates in very nearly the
same flavor basis as are the quarks \ref\fcphen{F. Gabbiani and A. Masiero,
\np{322}{1989}{235}}, \ref\nssquark{Y. Nir and N. Seiberg,
\pl{309}{1993}{337}}.  To discuss these constraints we use the notation and
analysis from \nssquark.

 The $6\times 6$ squark mass-squared matrix may be written as
\eqn\sqmmat{{\tilde M}^{q2} = \(
\matrix{{\tilde M}^{q2}_{LL} &{\tilde M}^{q2}_{LR} \cr
{\tilde M}^{q2\dagger}_{LR} &{\tilde M}^{q2}_{RR} \cr}\)}
where $L$ and $R$ refer to the chirality of the associated quarks.
Assuming that the $SU(2)\times U(1)$ violating $LR$ components of ${\tilde
M}^{q2}$  are smaller than the diagonal components, then
FCNC experiments limit the quantities
\eqn\deldef{\delta^q_{AB} = {V^q_A {\tilde M}^{q2}_{AB} V^{q\dagger}_B\over
\tilde m^2}\ ,}
where $V^u_{L,R}$ and $V^d_{L,R}$ are the
unitary matrices which diagonalize the $u$ and $d$ quark mass matrices.
The $\[\delta^d_{AB}\]_{12}$'s are constrained to be less than
$ few \times 10^{-3}$, while the  $\[\delta^d_{AB}\]_{13}$'s and
$\[\delta^u_{AB}\]_{12}$'s are constrained to be smaller than $few \times
10^{-2}$.  Various explanations of how these small numbers arise naturally have
been proposed, such as squark universality  and horizontal flavor symmetries.
Universality, as invoked in minimal supergravity \ref\hwein{L. Hall, J. Lykken
and S. Weinberg, \physrev{D27}{1983}{2359}}\ is quite unnatural, since there is
no reason why the physics that gives diverse Yukawa couplings to the different
families wouldn't also give diverse squark masses, but models have been
proposed where squark universality is a natural consequence of their identical
gauge interactions
\ref\squn{M. Dine and A.E. Nelson, \physrev{D48}{1993}{1277}}.  Explanations
for small FCNC based on horizontal symmetries \nref\horiz{M. Dine, A. Kagan and
R. Leigh, SLAC-PUB-6147, hep-ph-9304299 (submitted to Phys. Lett.
B)}\refs{\nssquark,\horiz} simply ensure that the inevitable breaking of flavor
symmetry in the squark sector is small enough for symmetry reasons to not have
been observed.  The model we are describing here falls into this second
category.

Our $\Delta(75)$ model has small FCNC effects due to the nonabelian flavor
symmetry, so long as the order parameter for SUSY breaking is flavor neutral.
First consider the $LR$ sector of the squark mass matrix.  One contribution is
proportional to the Yukawa coupling and is diagonal in the quark mass
eigenstate basis.  The other contribution arises through the soft SUSY
violating trilinear couplings of the squarks to the Higgs doublets.  These
couplings are assumed to arise from a dimension five superpotential $W'\sim
W\times \phi/M_p$, where $\phi$ is a chiral superfield whose $F$ component
breaks supersymmetry at an intermediate scale, and \hbox{``$\sim$''}  means
that there is a one-to-one correspondence between operators, although the
$\CO(1)$ coupling constants are not assumed to be the same.  This implies that
at low energy the effective trilinear couplings are
\eqn\tril{\tilde m\[\tilde Y_u \tilde Q H_u \tilde u^c +\tilde  Y_d\tilde Q H_d
\tilde d^c\]}
where the $\tilde Y$ matrices have the same texture as the Yukawa coupling
matrices.  Thus in the flavor basis where the quark masses are diagonal, the
$\{ij\}$ component of  ${\tilde M}^{q2}_{LR}$ is at most of order $\tilde m
\sqrt{m_im_j}$, where $m_i$ are the corresponding quark masses, and so their
contributions to
the constrained parameters $\delta^q_{LR}$ are very small.

The $LL$ and $RR$ parts of the squark mass matrix also get two contributions.
The first is proportional to $Y^{\dagger}Y$ and is diagonal in the quark mass
eigenstate basis.  The second arises from the dimension six $D$-terms,
\eqn\dsix{
{\phi^*\phi\over M_p^2}\[ c_1 \CF^* \CF + c_2\psi^*\psi + c_3
\bar\psi^*\bar\psi + \ldots\]_D\ .}
where the $\Delta(75)$ symmetry dictates that there is universality in the
coupling of the three families.  These terms alone give contributions to  the
$LL$ and $RR$ components of ${\tilde M}^{q2}$ which are proportional to the
unit matrix and hence diagonal in any basis.   FCNC effects can exist in
dimension eight operators arising directly from the Planck scale
\eqn\dseven{\Biggl. {\phi^*\phi S^* (\CF^* X\CF)\over M_p^4 }\Biggr|_D}
inducing off-diagonal contributions to $\delta^q_{LL,RR}$ of order
$\vev{S}M_{GUT}/M^2_p\simeq 2\times 10^{-5}$.
Larger contributions arise from  dimension eight operators generated by
integrating out the heavy $\psi$ field as in Fig. 4,
leading to the operator
\eqn\dsevenii{\Biggl.{\phi^*\phi(\CF^* XX^*\CF)\over M_p^2 \vev {S}^2
}\Biggr|_D\ .}
\topinsert
\centerline{\epsfbox{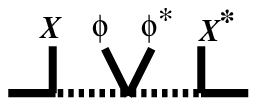}}
\caption{{\bf Fig. 4.} {\it Supergraph contributing to flavor changing squark
masses. The dotted line is the $\psi/\bar\psi$ field, and $\phi$ is the field
giving rise to supersymmetry breaking.}
  }
\endinsert\noindent
Since $\vev{ X/S}\equiv x\simeq  1/20$, this operator would appear to
contribute to FCNC at the $3\times 10^{-3}$ level.  However, $\vev{X^* X }$ in
the above operator
is flavor  diagonal in the $\Delta(75)$  basis we have been using, and
therefore gives rise to off-diagonal contributions in $\delta^q_{LL,RR}$  of
order $x^2\times \theta$, where $\theta$ is the relevant mixing angle.  In the
kaon system, for example,  this gives $\delta^d\simeq \theta_c/400 = 5\times
10^{-4}$. Thus FCNC in a model such as this one are below current limits, but
only by about an order of magnitude, even though flavor physics occurs up at
the GUT scale.

It is interesting to note that FCNC effects increase in supersymmetric models
as the flavor symmetry breaking scale gets closer to the Planck scale.  Thus it
is conceivable that improved searches for FCNC could in fact probe physics in
the region between the GUT and Planck scales.  This is peculiar to models such
as supersymmetry in which GIM violation can proceed through soft operators.

\newsec{Conclusions}
In this paper we are advocating using nonabelian discrete flavor symmetries for
unifying flavor at short distances.  The example we have given --- a
supersymmetric GUT with a  $\Delta(75)$ flavor symmetry --- can account for the
diversity of quark and lepton masses and mixings without small fundamental
parameters, other than the hierarchy of the mass scales $M_p$, $M_{GUT}$ and an
intermediate scale associated with the masses of vectorlike families.  This
particular model  predicts mixing angles to be approximately equal to their
observed values, as well as $\tan\beta\simeq 3$. The model also predicts a
seesaw mechanism for neutrino masses, with the  $\tau$ neutrino mass given
approximately by $M_p/ G_F M_{GUT}^2 \simeq 10$ eV. The two lighter neutrino
masses scale like the up-type quark masses squared (at the GUT scale) and are
much lighter.

We believe that our $\Delta(75)$ model exhibits a number of features that will
be generic in flavor unification models that do away with an explicit fermion
mass hierarchy put in by hand.  These include:
\item{(i)} Due to the extra families added in such schemes, the gauge group
$\beta$ function changes sign at short distances. This requires that flavor
symmetry breaking occur near the GUT scale or higher, or that there are larger
gauge groups at low energies than usually envisioned.  Typically, gauge
interactions are strong near $M_p$ in these models.   It is intriguing that a
model of flavor physics favors strongly interacting physics at the Planck
scale.
\item{(ii)} With flavor symmetry breaking occuring at a high scale, the light
quark masses and mixings are sensitive to operators suppressed by powers of
$M_p$.  In the model described here, the relatively large Cabibbo angle is due
to a dimension five operator.
\item{(iii)} Flavor changing neutral currents are typically suppressed enough
to be acceptable in such models, due to the nonabelian flavor symmetry.
However, the proximity of the flavor symmetry breaking scale to $M_p$ means
that  FCNC effects  from these ultrashort distance scales could be detectable.
\item{(iv)}  Due to supersymmetry, the most generic operators consistent with
flavor symmetry are {\it not} generated when heavy particles are integrated out
of the theory.  This suggests that an effective Lagrangian approach is no
substitute for a model of short distance flavor physics.

In models with the short distance flavor democracy we are advocating, Higgs
fields typically  carry family quantum numbers, and understanding symmetry
breaking becomes a more pressing issue.  An important problem sidestepped in
this paper has been the doublet-triplet splitting of the Higgs, which now
becomes entangled with the problem of flavor.  Other issues that remain to be
addressed in detail are neutrino masses and CP violation.

\vskip1in
\centerline{{\bf Acknowledgements}}

We wish to thank A. Cohen, A. Manohar, A. Nelson and N. Seiberg for useful
conversations. This work is supported in part by funds provided by the DOE
under contract \doe, by the NSF under contract \pyidk, by the Texas National
Research Laboratory Commission under contract \#RGFY93-206, and by the Alfred
Sloan Foundation.
\vfill\eject

\appendix{A}{Triplet decomposition in $\Delta(75)$}

Here we give the decomposition of the products of triplet representations shown
in table 2, consistent with the basis defined in eq. \basdef.  As discussed in
\S 2,  the generator $\hat E_{00}$  has the same representation matrix
$D_R(E_{00})$
for all of the triplet representations $R$:
\eqn\erep{D_R(E_{00}) =  \[\matrix{0&1&0 \cr 0&0 &1\cr 1&0&0\cr}\]\ ,\qquad
R=\{T_1,\ldots,\bar T_4\}\ .}
The representation matrices corresponding to the generator $\hat A_{10}$ are
given by $D_1(\hat A_{10}) = A_{10}$ and
\eqn\arep{
D_2(\hat A_{10}) = A_{20}\ ,\quad
D_3(\hat A_{10}) = A_{13}\ ,\quad
D_4(\hat A_{10}) = A_{21}\ ,}
where $D_n$ is the representation matrix for the triplet $T_n$ and the $A_{pq}$
matrices are defined in eq. \agen.  The above representations follow from the
conventions \basdef.  This is enough information to determine all of the
invariant tensors of the group.

{}From table 2 one sees that $T_n\otimes \bar T_n$ always contains all three
singlet representations, for $n=1,\dots,4$. Writing $T_n$ as $\{x,y,z\}$, one
finds these singlets to be
\eqn\sing{\eqalign{
T_n\otimes \bar T_n \vert_{A_1} &= x \bar x + y\bar y + z \bar z\cr
T_n\otimes \bar T_n \vert_{A_2} &= x \bar x + \omega y\bar y + \omega^2 z \bar
z\cr
T_n\otimes \bar T_n \vert_{\bar A_2} &= x \bar x + \omega^2 y\bar y + \omega z
\bar z\cr}}
where $\omega \equiv e^{2i\pi/3}$.

For the decomposition of a product of two triplets into a third triplet, it
suffices to give the structure of all of the three-triplet invariants.
Due to eq. \erep,  all invariants of three triplets $(ABC)$ can be specified by
three numbers $\{ijk\}$ signifying that $(ABC)= A_iB_jC_k + c.p.$, where $c.p.$
stands for cyclic permutation of each representation's index.  For example,
$(ABC) = \{112\}$
denotes that  $(A_1B_1C_2 + A_2 B_2 C_3 + A_3 B_3 C_1)$ is a $\Delta(75)$
singlet.  Table 2 reveals that the product of three triplets of a given
representation always contains two invariants.  These are given by
\eqn\tniii{(T_n T_n T_n)= \{123\} + \{213\}\ .}
Thus, for example, if one wants to find the $\bar T_1$'s contained in
$T_1\otimes T_1$, one finds them to be
\eqn\iiib{ T_1\otimes T'_1 \vert_{\bar T_1} = \trip{y z'}{zx'}{xy'},\ \trip{z
y'}{x z'}{ yx' }\ ,}
or any linear combination of the two.
There remain sixteen independent invariants with three triplets, and their
structure is found to be
\eqn\tripiii{\eqalign{
\{111\}:\ &(1 1 \bar 2)\, (122),\, (334),\, (344)\cr
\{112\}:\ &(1\bar 3 2),\, (1 4 \bar 3),\, (\bar 2 3 4),\,(\bar  2 4 1) \cr
\{113\}:\ &(1 3 2),\, (1 \bar 4 3),\, (2 3 4),\, (2 4 \bar 1) \cr
\{123\}:\ &(3 \bar 1 1),\, (\bar  4 1 4),\, (4 \bar  2 2),\, (2 \bar  3 3)\
.\cr}}
Thus for example, if one wanted to find the invariant formed from
$ \bar T_2 \otimes T_4 \otimes T_1$
one notes that $(\bar T_2 T_4 T_1)$ is an invariant of the $\{112\}$ type, so
that
\eqn\exampi{ \bar T_2 \otimes T_4 \otimes T_1\vert_{A_1} = \bar a \alpha y +
\bar b \beta z + \bar c \gamma x\ ,}
where we have taken $T_1 = \{x,  y,z\}$,  $\bar T_2 = \{\bar a,\bar b,\bar c\}$
and $T_4 = \{\alpha, \beta,\gamma\}$.  Similarly, if one wanted to find the
$T_4$ contained in $\bar T_1\otimes T_2$, the same $\{112\}$ invariant yields
\eqn\examp{\bar T_1\otimes T_2\vert_{T_4} = \trip{\bar y a}{ \bar z b}{ \bar x
c}\ .}

\footatend\vfill\eject\immediate\closeout\rfile\writestoppt
\baselineskip=14pt\centerline{{\bf References}}\bigskip{\frenchspacing%
\parindent=20pt\escapechar=` \input refs.tmp\vfill\eject}\nonfrenchspacing
\vfil\eject

\vfil\eject

\bye